# PETROLEUM, COAL AND OTHER ORGANICS IN SPACE

Franco Cataldo[1,2(*)], D. A. García-Hernández[3,4], Arturo Manchado[3,4,5]

[1] Actinium Chemical Research srl, Via Casilina 1626A, 00133 Rome, Italy
[2] INAF-Osservatorio Astrofisico di Catania, Via S. Sofia 78, 95123 Catania, Italy
[3] Instituto de Astrofísica de Canarias (IAC), C/ Via Lactea s/n, E-38205 La Laguna, Tenerife, Spain
[4] Departamento de Astrofísica, Universidad de La Laguna (ULL), E-38206 La Laguna, Tenerife, Spain
[5] Consejo Superior de Investigaciones Científicas, Madrid, Spain

(*) Corresponding author; e-mail: franco.cataldo@fastwebnet.it

## Abstract

The petroleum and coal models of the unidentified infrared emissions (UIE), sometimes referred also as unidentified infrared bands (UIBs) has been reviewed mainly based on the work of the authors with the inclusion of unpublished results. It is shown that the petroleum and coal model of the UIE converges and merges quite well with the MAON (Mixed Aromatic Aliphatic Organic Nanoparticles) model of the UIE. It is shown that the thermal treatment of various substrates like PAHs, alkylated PAHs but also mixed aliphatic/olefinic substrates leads invariable to carbonaceous materials matching the infrared spectrum of anthracite coal or certain petroleum fractions. Thus, the experimental thermal processing (which under space conditions could be equivalent to the expected processing by shock waves or high energy radiation) of mixed aromatic/aliphatic organic matter can be used to match also the UIE evolution. Another way to simulate the thermal/radiation processing of organic matter in space, can be achieved through the carbon arc. Simple substrates processed in this way produce carbon soot and a plethora of organic molecules. Fullerenes are found in space both through mid-infrared and optical spectroscopy and it is very likely that other complex related species such as endohedral fullerenes (i.e. fullerenes with a metal, heteroatom or molecules inside the cage) may be formed in space. After all, their formation requires the same conditions as those needed for fullerene formation provided that also a metal vapour (e.g. interstellar/circumstellar gas) is available. The last part of this review is thus dedicated to the recent results on the study and characterization of an endohedral $C_{60}$ derivative containing lithium inside the cage.

**Key Words:** UIE; UIBs; coal model; petroleum fractions model; infrared spectroscopy; carbonization; carbon arc; endohedral fullerenes.

## 1 Introduction

The unidentified infrared emissions (UIE), sometimes reported also as unidentified infrared bands (UIBs), are quite ubiquitous in the cosmos, being detectable from a number of different astronomical objects, from galactic nuclei to the diffuse interstellar medium (ISM) and from young stellar objects to old evolved stars like protoplanetary and planetary nebulae. This subject has been reviewed by several authors, like for instance Kwok & Zhang (2011), Kwok (2012, 2016), Peeters (2013), Mackie et al. (2015) and references therein. The UIE phenomenon has puzzled astronomers and astrochemists for decades and a series of possible carriers of the UIE were proposed (Tielens 2008; Kwok, 2012, 2016, 2019; Mackie et al. 2015). In Table 1 (first column in the left) we list the most common and prominent UIE features, while weaker UIE features are listed in the second column from left of Table 1. The most popular carrier/s of UIE were thought to be polycyclic aromatic hydrocarbons (PAHs; see e.g. Tielens 2008, Peeters 2013 for a review). Galué & Leines (2017) have recently suggested that large and curved PAHs may play a key contribution to the UIE. However, as shown by Zhang & Kwok (2014), no single, individual and well defined PAH specie has yet been identified in any of the astronomical objects carrying the UIE. Perhaps, the only exception may be represented by the recent radioastronomical detection of benzonitrile towards the Taurus molecular cloud (McGuire et al. 2018). Consequently, the UIE have also been suggested to be the result of a mixture of different carriers and the individual molecules of these carriers should have a complex structure. In line with this view, it was certainly fecund and stimulating the proposal from Papoular et al. (1989), Papoular et al. (1996) as well as Guillois et al. (1996) that a model for the UIE could be coal and especially certain mature or anthracite coals could be used to match the UIE. Papoular (2001), later broadened the coal model into the kerogen model as a tool for understanding the evolution of the interstellar carbonaceous dust.

From the coal and from the kerogen model of the UIE, it was quite natural to discover that certain heavy petroleum fractions, i.e. certain distillation residues of crude petroleum display infrared spectra matching those of certain coals and moreover of certain proto-planetary nebulae as shown by Cataldo et al. (2002, 2003, 2004) and Cataldo, Keheyan & Baccaro (2004). These heavy petroleum fractions are made by a mixture of relatively large and complex molecules which can be described as mixed aliphatic-aromatic-naphthenic, containing also some heteroatoms as reviewed in some occasion by Cataldo et al. (2012, 2016). These views led Kwok & Zhang (2011, 2013) to develop the mixed aromatic-aliphatic organic nanoparticles (MAONS) model for the UIE, suggesting that solid and amorphous nanoparticles with complex chemical structures, rather than gas-phase and free-flying PAH molecules are the carriers of the UIE. Through spectral fittings of the astronomical spectra of the UIE, Kwok & Zhang (2011) have shown that a significant amount of the energy is emitted by the aliphatic component, implying that aliphatic groups are an essential part of the chemical structure. The MAON concept was also explored further by Papoular (2013, 2014). In the context of the petroleum fractions, Cataldo, García-Hernández & Manchado (2013) and Cataldo et al. (2013) have shown that examples of naturally occurring MAONs can be represented by petroleum asphaltenes as well as asphaltenes derived by other fossil hydrocarbons. The asphaltenes are made by a series of condensed aromatic rings together with naphthenic rings. The rings are substituted by short and long alkyl chains both linear and branched and, some of the carbon atoms of the structure is substituted by heteroatoms, essentially O, N, and S. Examples of possible asphaltene structures can be found in Cataldo, García-Hernández & Manchado (2013) and Cataldo et al. (2013) With the occasion of the present work, a series of unpublished data regarding the heavy petroleum fractions, selected asphaltenes and carbonized molecules (Cataldo, García-Hernández & Manchado 2018), showing spectra similar to that of coal and UIE, are presented and discussed.

## 2. Experimental

### 2.1 - Materials

A sample of standard anthracite coal (reference material no. 65) was supplied by the Commission of the European Communities 'Community Bureau of Reference BCR' with an individual identification analysis no. 417. RAE (Residual aromatic extract) heavy oil fraction was a standard commercial sample produced by a well-known multinational oil company. All the PAHs samples, naphthalene, acenaphthylene, anthracene, tetracene, pentacene, decacyclene and hexamethylnaphthalene were purchased from Aldrich-Merck. Also the sesquiterpene guaiazulene was obtained from Aldrich-Merck. A sample of $[Li@C_{60}]^+PF_6^-$ was purchased from Idea International Co. Sendai, Japan. The spectroscopy matrices cesium iodide, low molecular weight polyethylene (PE), potassium bromide and adamantane where obtained from Aldrich-Merck. The reference sample of $C_{60}$ 99+% purity was obtained from MTR Ltd, USA.

### 2.2 – Laboratory equipment

The mid-infrared spectra were recorded in transmittance on Nicolet 6700 FT-IR spectrometer from Thermo-Scientific. The low-temperature apparatus consisted of a variable temperature cell from Specac model P/N 21525 equipped with heated KBr windows and a temperature-controlled sample holder, which is able to work in the range between +250°C to -180°C. The low temperature limit was reached using liquid nitrogen. The variable temperature cell was evacuated with a Buchi vacuum pump model V-710 equipped with four diaphragm heads and a three-stage vacuum creation process, which delivers 3.1 $m^3$ $h^{-1}$ and an absolute vacuum of 2 mbar. The far-infrared spectra from 600 $cm^{-1}$ down to 50 $cm^{-1}$ were recorded with the Nicolet 6700 FT-IR spectrometer after opportune adjustments of the optical bench (change of the beam splitter and the detector). Furthermore, the entire spectrometer optical bench and the sample compartment was continuously purged with completely dry air free from $CO_2$ at 12 L/min. The CsI, KBr or adamantane pellet used in the infrared spectra measurement was produced by a Silfradent double column press equipped with a pressure meter. The thickness and the diameter of the pellets was measured with a Somet digital micrometer having a sensitivity of 0.01 mm.

## 3. Results and Discussion

### 3.1 – Infrared spectroscopy of anthracite coal in the range between +250°C and -180°C

By using a similar experimental set-up, the infrared spectra of fullerenes and fulleranes were previously studied in a wide range of temperatures, i.e. from +250°C to -180°C by Iglesias-Groth et al. (2011, 2012) and by Cataldo et al. (2012, 2013). The infrared spectra collected at different temperatures, permit to know exactly what are the temperature sensitive infrared bands and to follow their shift in band positions as function of temperature. For instance, some of such temperature-sensitive bands were identified in the spectra of $C_{60}$ and $C_{70}$ fullerenes by Iglesias-Groth et al. (2011, 2012) and their temperature shift was extrapolated down to 0 K. Another advantage of low temperature spectroscopy, regards the fact that the infrared bands show a sharper shape and a minimized broadness. This is a consequence of quantomechanical effects on the accessible density of states for each vibrational transition (Avram et al. 1972). Furthermore, some buried absorption bands not easily detectable for example at room temperature or at higher temperatures are easily detectable at very low temperature. To the best of our knowledge such temperature spectra in a wide range of temperature were never reported on anthracite coal. Thus, with the hope to obtain a better defined infrared spectrum on anthracite coal, the spectra were recorded under high vacuum both at -180°C (i.e. at liquid nitrogen temperature) as well as at +250°C, a temperature at which coal is still stable. The mentioned coal spectra at the two extreme temperatures are shown in Fig. 1a and 1b. Furthermore, the infrared bands are also listed in Table 1. Unfortunately, the very complex chemical nature of anthracite coal does not give the spectacular results obtained in the case of $C_{60}$ and $C_{70}$ low temperature infrared spectra in terms low temperature band shifting and sharping. However, Fig. 1a shows that the infrared band at 6.95 μm recorded at -180°C is much sharper and "cleaner" than the corresponding band at 6.94 μm observed at +150°C. The same comments apply also to the bands at 7.29 and 9.67 μm recorded at -180°C with respect to the same bands at +250°C. Even more eloquent are some features in the spectra of Fig. 1b. For example, the band at 11.60 μm recorded at +250°C appears as a doublet respectively at 11.53 and 11.85 μm when observed at -180°C. The shoulders at 12.04 μm and another at 13.21 μm are observable in the left side of the bands at 12.42 and 13.40 μm respectively, if recorded at -180°C. Such shoulders are not detected in the spectrum obtained at +250°C. These results show very clearly that the improvements in the definition of the infrared bands of coal recorded at -180°C are evident with respect to the spectrum recorded at +250°C but not as decisive and spectacular as in the case of infrared spectra recorded on pure substances. Furthermore, the infrared data collected in Table 1, suggest that there is a general trend to the shift toward shorter wavelengths (with some exceptions) for the infrared coal bands recorded at -180°C in comparison to the same bands recorded at +250°C. However, the observed shift is really minimal in comparison to the shift observed on fullerenes infrared spectra at extreme temperatures (Iglesias-Groth et al. 2011). The infrared band shift toward shorter wavelengths at lower temperature is known as "temperature effect" is due to a combination of different effects, involving the already mentioned quantomechanical density of states and changes in bond length with temperature (and hence force constant) (Smith 1956; Roy 1968; Avram et al.1972). The minimal infrared band shift observed on anthracite coal spectrum at -180°C, is again explainable with the complex and macromolecular chemical structure of coal as testified by the numerous molecular models proposed to describe it and consisting indeed in crosslinked mixed aromatic-aliphatic and cycloaliphatic (or naphthenic) structures trapping smaller molecules in a host-guest interaction as shown by Mathews et al. (2011), Mathews & Chaffe (2012) and Qin (2018).

Furthermore, in agreement with the MAON model, anthracite coal is composed by elemental carbon and hydrogen, i.e. C = 92.0%, H= 4.1% but also by O = 2.2%, N = 1.2% and S = 0.5 %. As suggested by Van Krevelen (1993), the elemental composition reported may vary within a certain range, but heteroatoms are always present in anthracite coal. Another interesting property of the coal model connected to the MAON one regards the fact that X-ray diffraction studies have revealed that in high rank coal the graphene sheets (called also aromatic lamellae) have a diameter of approximately 300 nm and about 4 or 5 are stacked each other forming some kind of aromatic clusters (Van Krevelen 1993) which can be viewed as the core of MAONs.

## 3.2 – Infrared spectroscopy of petroleum fraction RAE at +50°C and -180°C

As already discussed in the introduction, certain aromatic petroleum fractions are valid alternatives to anthracite coal. Recently, a freshly prepared heavy petroleum fraction was studied by FT-IR at the liquid nitrogen temperature. The heavy petroleum fraction studied is called "RAE" i.e. a residual aromatic extract. After the crude oil atmospheric and vacuum distillation, the bottom residue is extracted with a solvent to remove the asphalt. Afterwards the deasphaltized fraction is extracted again with another solvent. The residue of this extraction (i.e. the insoluble fraction) is in fact called RAE or residual aromatic extract while the extractable fraction is a bright stock base oil which is sent to a dewaxing unit and is then used for the preparation of lubricants and/or lubricants additives. Regarding the RAE composition, chromatographic analysis has revealed that it is composed by aromatics by 40% by weight, while paraffinics and naphthenics represent the remaining 60% by weight (about 30% each).

The typical sulphur content of RAE is 3.5% by weight and total heteroatoms content intended as O+N+S is about 5%. RAE is a highly viscous oil at room temperature with a viscosity at 50°C of 1730 $mm^2$/s with a refractive index of 1.5575 measured at 20°C. Although the average hydrocarbons present in RAE can be described as mixed aromatic, paraffinic and naphthenic hydrocarbons, fulfilling the requirements of the MAONs model, the mass spectrometric analysis reveals also the presence of small amounts of free PAHs. For example, as shown in Scheme 1, a series of fluoranthene, anthracene and pyrene derivatives can be detected although each individual PAH rarely exceeds 1 ppm concentration in the RAE mass.

Fig. 2 shows the infrared spectrum of RAE recorded both at +50°C and also at -180°C. As expected at the liquid nitrogen temperature the infrared bands of RAE appear sharper, more defined and also a few weaker bands not detectable at +50°C can be observed. In Table 1 all the infrared bands of RAE are also listed. The mixed aromatic/aliphatic/naphthenic nature of RAE is confirmed by the infrared bands assigned to one of these three categories, although the predominance of the aromatic component is confirmed also by infrared spectroscopy.

The assignments of the most important infrared bands are reported in Table 1 together with the most prominent and weaker UIE features. Fig. 2 shows that the out-of-plane bending region of RAE is richer in absorption bands than the case of anthracite coal and this may imply a richer variety of different types of molecules contributing in this spectral region. Previously, Cataldo, García-Hernández & Manchado (2013) as well as Cataldo et al. (2013) have shown both with mid- and far-infrared spectroscopy that the core structure of the RAE petroleum fraction could be viewed as a highly substituted tetrahydronaphthalene and/or 9,10-dihydroanthracene. Furthermore, in Cataldo, García-Hernández & Manchado (2013) it was speculated there exists a possible connection between the still unidentified 21 μm feature and the hydrogenated acene structure of the core structure of selected petroleum fractions. The carrier of the 21 μm feature was extensively discussed by Kwok (2012; 2016) arguing that being the 21 μm feature found mainly in carbon-rich astrophysical objects it is highly probable that it is organic chemical nature. Moreover, according to Kwok (2012; 2016), the 21 μm feature seems to be related to another band at 30 μm and the two bands could be due to the same carrier. However, astrophysical observations on a series of astrophysical objects suggest that the carrier of these two bands is synthesized in the circumstellar environment after the AGB stage (Kwok, 2012; 2016). Molecules which are thermodynamically stable at high temperatures like for example fullerenes, polyynes and so forth, could be among the possible candidates as carrier of the 21 μm feature.

### 3.3 – Carbonization of selected substrates to match the anthracite coal spectrum

The seduction of the coal model of UIE and the related petroleum fraction and asphaltene models which at the end can be collected into the more general MAONs model regards the fact that it does not exist a single reference spectrum but samples of both coal, petroleum fractions and asphaltenes can be selected at different degrees of "maturity" or ranks. It is possible, for instance, to start from a low rank coal that is richer in hydrogen content and aliphatics and by a thermal or radiation treatment reach a higher rank coal with lower hydrogen content and higher aromatic content (Papoular, 2013, 2014; Cataldo & Keheyan, 2003; Cataldo, Keheyan & Baccaro, 2004). In parallel, also the organic matter in space - for example, as found around evolved stars such as asymptotic giant branch (AGB) stars and (proto-)planetary nebulae - responsible for the UIE undergoes a considerable evolution since its mixed aliphatic/aromatic ratio changes continuously, affecting the resulting UIE features. Indeed, the UIE, as shown e.g. by Kwok (2012), display a different emission band pattern in several space environments depending on the degree of processing (e.g., by UV radiation, shocks, etc.) undergone by the organic matter. In this context, Cataldo, García-Hernández & Manchado (2018) started systematic thermal carbonization studies of selected acenes, showing that, at the end of the thermal treatment, the infrared spectrum obtained is very similar to the one of a high rank coal. In Fig. 3 we display the infrared spectra of the pristine acenes; namely naphthalene, anthracene, tetracene and pentacene. Furthermore, Fig. 3 shows also the infrared spectra of acenaphthylene, which is not an acene *stricto sensu* (see its structure in Scheme 2), as well as the spectra of the PAHs decacyclene and heptamethylnaphthalene (see also Scheme 2). While the carbonization studies of all acenes were already reported by Cataldo, García-Hernández & Manchado (2018), in this review the additional available results concerning the carbonization of acenaphthylene, decacyclene and heptamethylnaphthalene are reported for the first time. The former two PAHs are interesting since both are characterized in their structure by the presence of pentagonal rings, recalling the fullerene-like structure (Cataldo, 2002). On the other hand, heptamethylnaphthalene was selected because of its mixed aromatic/aliphatic character. Finally, also the naturally occurring sesquiterpene guaiazulene (1,4-dimethyl-7-isopropyl-azulene) was selected for the carbonization study; essentially because it has a lower aromatic character than the corresponding dimethyl-isopropyl-naphthalene and because of its mixed aliphatic/aromatic character.

When the acenes are heated in a sealed crucible under inert atmosphere to 700°C, a carbonization reaction occurs, changing the acene into a coal-like carbonaceous material as discussed by Cataldo, García-Hernández & Manchado (2018). The result of such a thermal treatment is thus the production of a material that basically displays the same infrared spectrum of anthracite coal, as shown in Fig. 4. Under our experimental conditions the carbonization of acenes was achieved by a thermal treatment, but as discussed in Cataldo, García-Hernández & Manchado (2018) similar results would be expected in some astronomical environments; e.g. under the action of shock waves or high intensity radiation fields (i.e. conditions existing in circumstellar environments).

Acenaphthylene is a naphthalene derivative with an additional pentagonal ring containing a double bond (Scheme 2). Consequently, acenaphthylene can be considered as a compound with mixed aromatic/olefinic character. In Fig. 3, the pristine acenaphthylene spectrum is dominated by three C-H bending features at 12.05, 12.90 and 13.73 μm. While the first two bending vibrations could be attributed to the three adjacent hydrogen atoms of the naphthalene ring, the latter infrared band can be assigned to the cis C-H wagging of the olefinic double bond (Colthup et al.1990).

As shown in Fig. 4, the thermal treatment at 700°C turns the acenaphthylene compound into a coal-like material, displaying basically the same band pattern of anthracite coal as well as the one for the other carbonized acenes. Despite the mixed aromatic/olefinic character, the final carbonization product of acenaphthylene is very similar to the carbonization product of the acenes. The same comments are applicable to both guaiazulene (characterized by a lower aromatic character than naphthalene and by three alkyl substituents; see Scheme 2) and to heptamethylnapthalene (which is a highly alkyl-substituted naphthalene with a more pronounced aliphatic character than its aromatic counterpart). Again, despite these unique structural features of guaiazulene and heptamethylnaphthalene (see the pristine spectra in Fig. 3), their carbonization at 700°C leads invariably toward the same spectrum of anthracite coal, as shown in Fig. 4.

The lesson we may learn from these experimental results is that, apparently, it does not matter too much how complex the starting MAONs could be (i.e. no matter too much about the specific aliphatic/olefinic/naphtenic and aromatic character) and the experimental thermal treatment (which under space conditions could be equivalent to the expected processing by high energy radiation or shock waves) leads invariably to a coal-like structure, matching in part (or completely) the UIE. Table 2 provides a synopsis of the infrared bands of the carbonized substrates treated in this section and the UIE.

## 3.4 – Carbonization of selected substrates in a carbon arc

Another topic closely related to the previous section, is the carbonization of selected substrates in a carbon arc. We remind that the main products of the carbon arc are polyynes, monocyanopolyynes and dicyanopolyynes, depending from the arcing conditions. Under helium atmosphere, however, the carbon arc produces fullerenes. This subject was fully reviewed quite recently by Cataldo, García-Hernández, Manchado, & Kwok (2016). Earlier works on the carbon arc products and possible implications for astrochemistry can be found, for example, in Cataldo (2004, 2005, 2006a,b). The carbon arc involves high strength electromagnetic fields and highly accelerated electrons and ions in a restricted volume, which permits to study the chemistry of selected substrates under extreme conditions involving temperatures of thousands of Kelvin. Recently, Cataldo, García-Hernández & Manchado (2017, 2019a) focused on the plasmalysis of simple molecules like benzene and toluene and the products were accurately identified by mass spectrometry. More than 73 different molecules were identified in the products from the benzene carbon arc. The main products, other than abundant carbon soot, were phenylacetylene, biphenyl, diethynylbenzene and the polyyne octatetrayne $H-(C\equiv C)_4-H$. Other products detected by GC-MS were styrene, indene, biphenylene, diphenylacetylene, naphthalene, acenaphtylene and even azulene, together with a rich mixture of PAHs. The key intermediate species in the toluene pyrolysis reaction is the formation of the benzyl radical accompanied by lower amounts of the phenyl radical and other radicals. The recombination of these radicals and the reaction with neutrals yields a plethora of products. About 72 different molecular species were detected from the arc pyrolysis of toluene. This time, the most abundant species was found bibenzyl, as derived from the recombination of the benzyl radicals followed by naphthalene and biphenylene. Interestingly, the benzyl radical may play a role in astrochemistry (Dangi et al. 2014) and the hydrocarbon processing with the arc provides access to extreme conditions simulating those present in certain interstellar and circumstellar media (i.e. conditions involving high temperatures and high energy conditions; see e.g. Keheyan et al. 2004; Yeghikian, 2011; Micelotta et al. 2010a,b; 2011). Curiously, polyynes, which are known to be quite abundant in the circumstellar envelopes of evolved stars like proto-planetary nebulae (see e.g. Cernicharo et al. 2001), were found much less abundant in the carbon arc pyrolysis of toluene in comparison with the arc pyrolysis of benzene.

## 3.5 – Endohedral fullerenes

Fullerenes were first detected in space as neutral molecules through mid-infrared spectroscopy in young planetary nebulae and reflection nebulae (Cami et al. 2010; García-Hernández et al. 2010,2011; Sellgren et al. 2010). Later, neutral fullerenes were detected in a variety of space environments (e.g. García-Hernández, Rao & Lambert 2011; Roberts, Smith & Sarre 2012, among others) and, more recently, ionized fullerenes such as the cation $C_{60}^+$ have been detected in the interstellar medium through the electronic transitions in the optical domain (Campbell et al. 2015). Due to the high reactivity of fullerenes with atomic hydrogen (Cataldo & Iglesias-Groth 2010; Iglesias-Groth et. al. 2012), the resulting hydrogenated fullerenes were searched in different space environments (Díaz-Luis et al. 2016; Zhang et al. 2017) with inconclusive results, but suggesting the hydrogenated fullerenes with very low H content as the promising candidates to be found in space. Fullerenes may also form adducts with different molecules in the external part of their cage and this stimulated the search for certain fullerenes Diels-Alder adducts with PAHs (García-Hernández, Cataldo & Manchado 2013) as well as with certain organometallic molecules (García-Hernández, Cataldo & Manchado 2016). Such fullerene adducts display spectral features very similar to those from the isolated fullerenes and if they form under astrophysical conditions (being stable/abundant enough), they may then contribute to the infrared emission features observed in fullerene-rich circumstellar/interstellar environments.

Remarkably, fullerenes form also endohedral structures with the incorporation inside the cage of metal and non-metal species and even small molecular clusters. Essentially, it is possible to distinguish two classes of endohedral fullerenes the EMFs and the NMEFs. The endohedral metallofullerenes (EMFs) consist of fullerene trapping inside the cage 1, 2, 3 metal atoms or clusters of up to 4 metal atoms including also non-metals such as C,N,O and S. Instead, the non-metal endohedral fullerenes (NMEFs) encapsulate noble gases, non-metal atoms like N and P or molecules like $H_2$, $N_2$, CO and $H_2O$. This subject was recently reviewed by Cataldo, García-Hernández & Manchado (2019b). The most surprising point regarding the formation of endohedral fullerenes regards the fact that especially EMFs are formed in a carbon plasma, in the presence of a buffer gas (e.g. helium but also nitrogen, ammonia…) and moreover in the presence of a metal vapor. In other words, wherever fullerenes are formed in space, if a metal vapor is present (e.g. interstellar/circumstellar gas), also endohedral fullerenes may be readily formed. In fact, the typical laboratory conditions for the synthesis of EMFs involve the use of a carbon arc, laser ablation or a radiofrequency furnace method. However, EMFs are not so easily accessible as $C_{60}$ and $C_{70}$ fullerenes. To date, the most commercially accessible EMF is $[Li@C_{60}]^+PF_6^-$. This is under the form of an ionic salt for stability and easily handling reasons. In the recent study made on $[Li@C_{60}]^+PF_6^-$ by Cataldo, García-Hernández & Manchado (2019b), it was found that it has a crystalline melting point with the onset at 86°C and peak at 90°C and a melting enthalpy (or heat of fusion) of 9.5 kJ/mol. The electronic absorption spectrum of $[Li@C_{60}]^+PF_6^-$ was found very similar to that of $C_{60}$. In the UV, however, all the electronic transitions of $[Li@C_{60}]^+PF_6^-$ show a small hypsochromic shift with respect to the corresponding transitions of $C_{60}$, thus requiring 5.2 kJ/mol more energy than the corresponding transitions in $C_{60}$ to occur. The incarcerated lithium cation causes alterations in the energy levels of the $C_{60}$ molecule leading to observed hypsochromic shift and hyperchromic effect in the molar extinction coefficients. The $[Li@C_{60}]^+$ displays in the mid-infrared the same and typical four band (at ~7.0, 8.5, 17.4 and 18.9 μm) pattern of neutral $C_{60}$, but the bands appear slightly shifted from their typical position. It is believed that the band shift is caused by the motion of the $Li^+$ cation inside the carbon cage. However, the presence of $Li^+$ inside the fullerene cage does not cause alterations in the geometry of $C_{60}$ and the truncated icosahedron geometry is fully preserved, at least at room temperature.

In the far-infrared $[Li@C_{60}]^+PF_6^-$ presents four extra bands that are not present in the far-infrared spectrum of neutral $C_{60}$. These bands were found at ~61.0, 65.8, 131.6 and 250.0 μm. The two far infrared bands at higher frequency are specifically due to the dynamics of the lithium cation inside the $C_{60}$ cage, while the band at 131.6 μm is due to the transverse optical phonon mode of the crystal $[Li@C_{60}]^+PF_6^-$ involving the interaction of $Li^+$ inside the carbon cage with the external $PF_6^-$ anion. The far-infrared band at 250.0 μm was assigned to the free rotation of the lithium cation inside the $C_{60}$ cage. Consequently, the search and identification of endohedral fullerenes in space should rely more on the far-infrared (terahertz) spectroscopy and the most diagnostic bands are those at 61.0, 65.8 and 250.0 μm, since the 131.6 μm is due to a solid state crystal structure phonon mode. Further studies on the infrared spectra of $[Li@C_{60}]^+PF_6^-$ in an astrochemical perspective were conducted by Cataldo, García-Hernández & Manchado (2020) in the temperature range comprised from -180°C to +250°C. It was found that the infrared band shift as function of temperature measured on $[Li@C_{60}]^+PF_6^-$ can be described by the same equations found on pure neutral $C_{60}$. Moreover, the integrated molar absorptivity ψ of each infrared band of $[Li@C_{60}]^+PF_6^-$ was found very close to the ψ value measured on pure neutral $C_{60}$. Only the molar extinction coefficients ε of the infrared bands were found systematically higher (but only by a factor 1.7≈1.8) on $[Li@C_{60}]^+PF_6^-$ with respect to the ε values measured on $C_{60}$. These results certainly imply that endohedral fullerenes could still contribute to the mid-IR features seen in fullerene-containing circumstellar/interstellar environments.

## Conclusions

The ubiquitous unidentified infrared emissions (UIE) widely detected in different astronomical environments and with different degrees of thermal/radiation processing could be matched and explained through the heavy petroleum fractions and/or the coal models. It has been shown that the MAONs model is able to summarize into a single and unified concept both the prerogatives of the petroleum fractions and the coal model. After all, the MAON model suggests that the carriers of the UIE are complex organic nanoparticles with mixed aromatic/aliphatic and naphthenic character, containing also heteroatoms such as O, N and S. Thermal and/or radiation processing of the MAONs can change the ratio aromatic/aliphatic/naphthenic, explaining also certain variability of the UIE as observed in different astrophysical objects. Since the MAON molecules are very complex, it is thought that the closer structures could be offered by those found in heavy petroleum fractions, asphaltenes and coal. Thermal and radiation processing of organic substrates more or less related to MAONs, can be performed on laboratory scale leading to the formation of carbonaceous products whose infrared spectrum is similar to that of coal or certain petroleum fractions.

Although fullerenes are firmly detected in space, it is thought that the next related species to be detected are fulleranes (the hydrogenated fullerenes) and endohedral fullerenes (i.e. fullerenes with atoms or molecules trapped inside the cage). The present review, is concluded with a review of the results available on the most accessible endohedral fullerene: $[Li@C_{60}]^{+}PF_{6}^{-}$.

## ACKNOWLEDGEMENTS

The authors acknowledge support from the State Research Agency (AEI) of the Spanish Ministry of Science, Innovation and Universities (MCIU) and the European Regional Development Fund (FEDER) under grant AYA2017-88254-P.

## ORCID

Franco Cataldo http://orcid.org/0000-0003-2607-4414

D. A. García-Hernández http://orcid.org/0000-0002-1693-2721

Arturo Manchado https://orcid.org/0000-0002-3011-686X

Figures

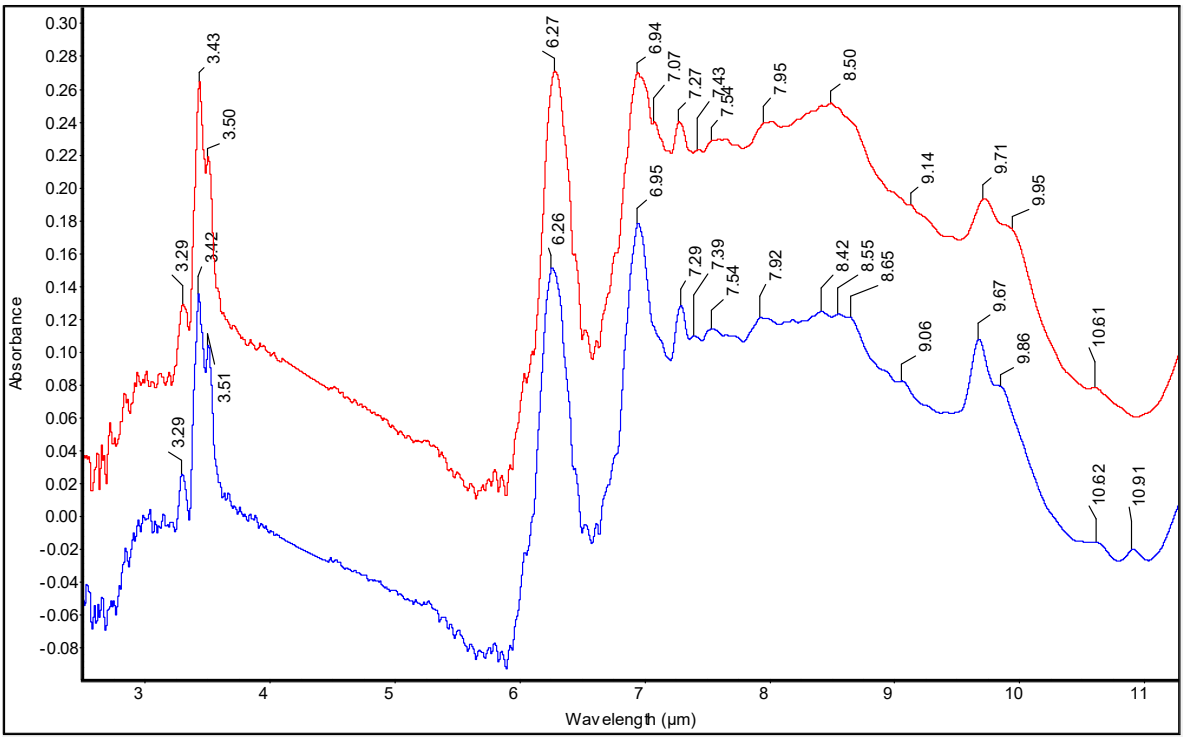

Fig. 1a – FT-IR spectra of anthracite coal (between 2.50 and 11.50 μm) embedded in KBr and recorded at +250°C (top spectrum red line) and at -180°C (bottom spectrum, blue line) in high vacuum conditions. The absorption bands are labelled in microns. The two spectra were slightly offset for clarity.

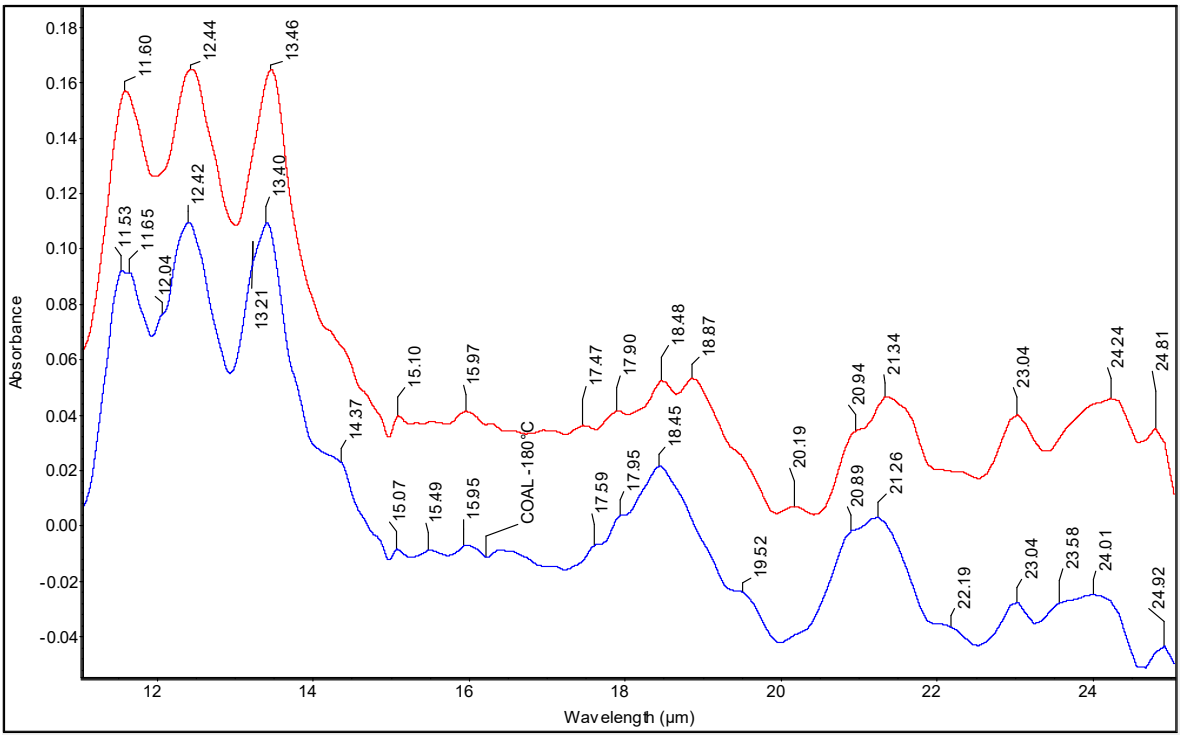

Fig. 1b – FT-IR spectra of anthracite coal (between 11.50 and 25.00 μm) embedded in KBr and recorded at +250°C (top spectrum red line) and at -180°C (bottom spectrum, blue line) in high vacuum conditions. The absorption bands are labelled in microns. The two spectra were slightly offset for clarity.

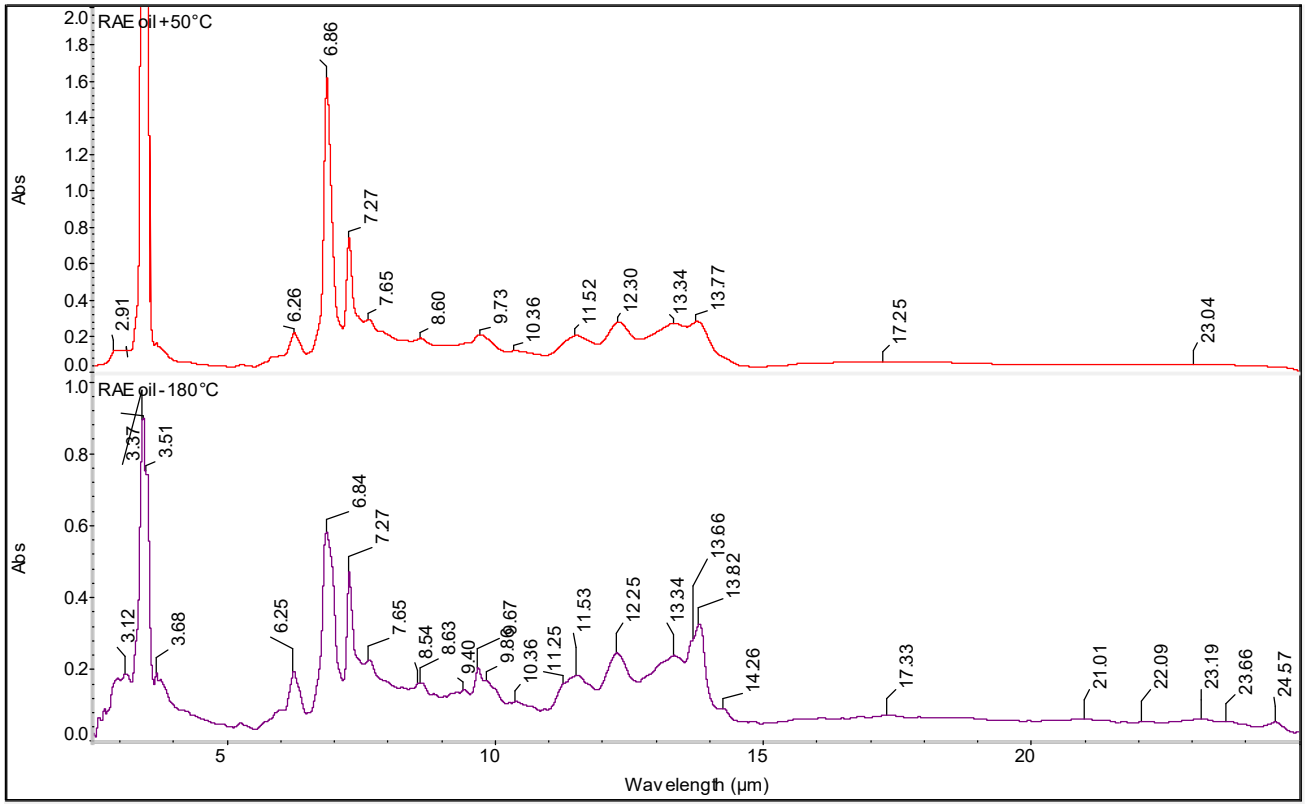

Fig. 2 – FT-IR spectra of the petroleum fraction "Residual Aromatic Extract" i.e. RAE thin film between CsI disks and recorded at +50°C (top spectrum) and at -180°C (bottom spectrum) in high vacuum conditions. The absorption bands are labelled in microns.

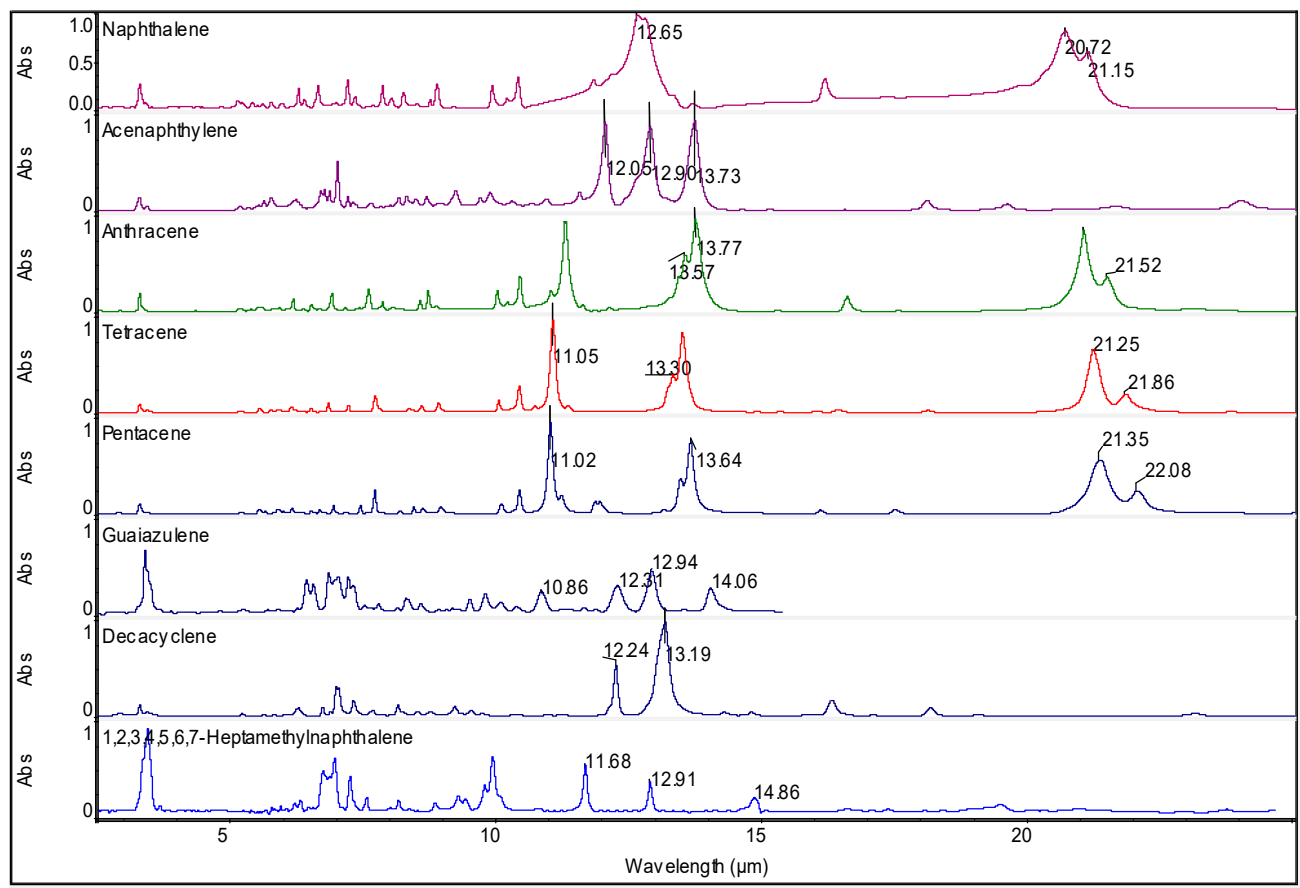

Fig. 3 – FT-IR spectra of pristine acenes (from top to bottom): naphthalene, acenaphthylene, anthracene, tetracene and pentacene. The spectra of sesquiterpene guaiazulene, decacyclene and heptamethylnaphthalene are shown in the bottom panels.

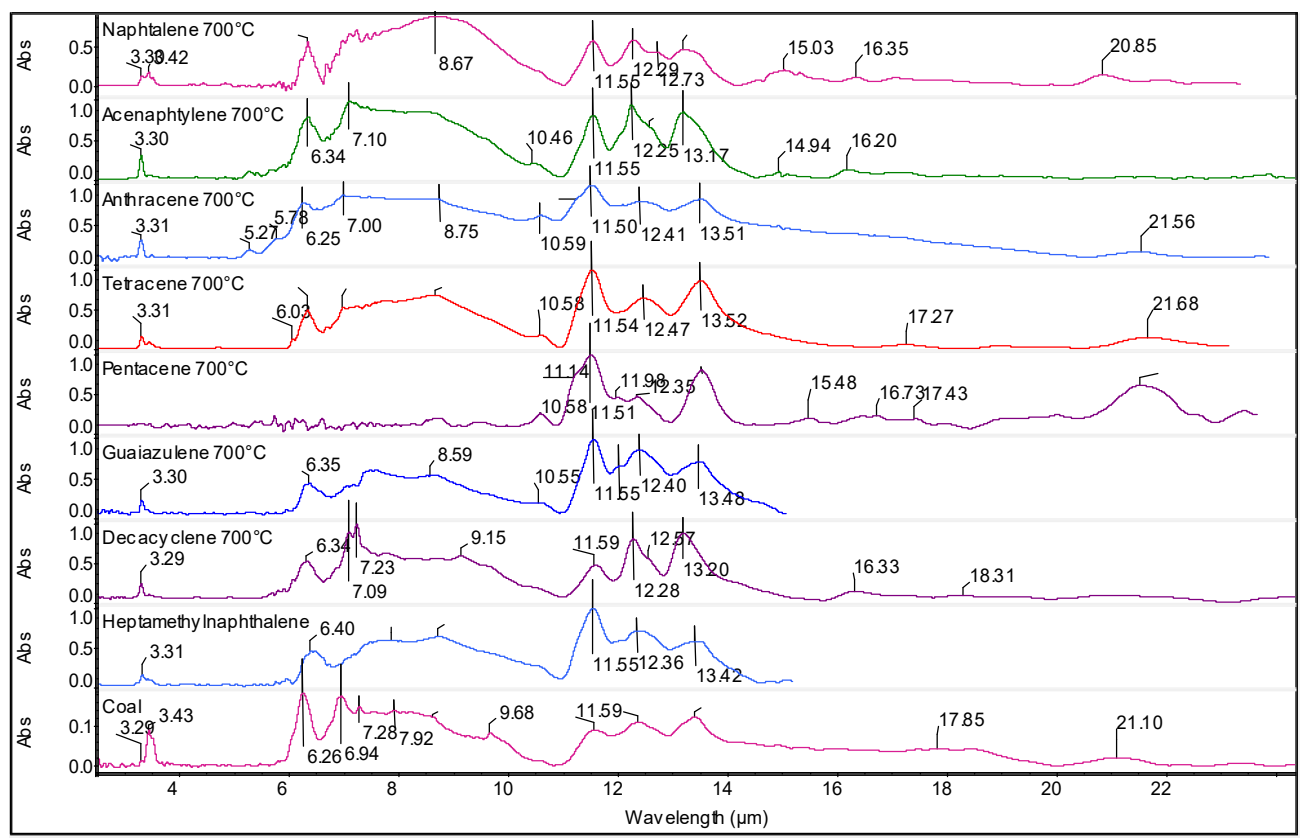

Fig. 4 – FT-IR spectra of carbonized acenes at 700°C under inert atmosphere. From top to bottom: carbonized naphthalene, acenaphtylene, anthracene, tetracene and pentacene. In addition to acenes, also the sesquiterpene guaiazulene (see Scheme 2), the PAHs decacyclene and the alkylated PAH heptamethylnaphthalene were carbonized. The reference spectrum of anthracite coal is shown at the bottom panel.

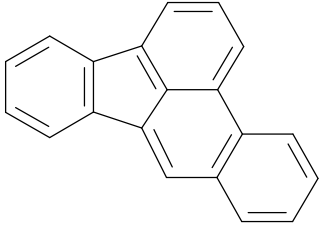

benzo[b]fluoranthene

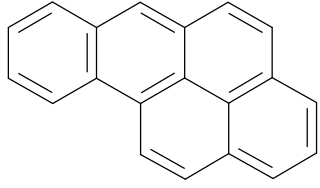

benzo[a]pyrene

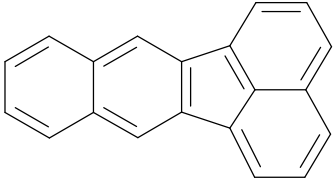

benzo[k]fluoranthene

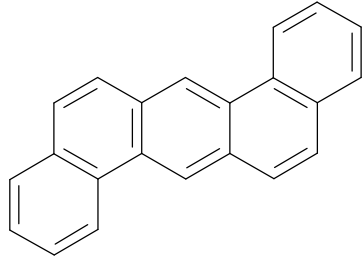

dibenzo[a,h]anthracene

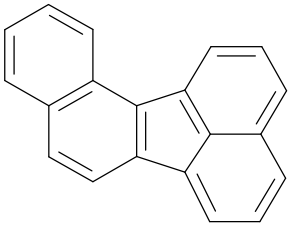

benzo[j]fluoranthene

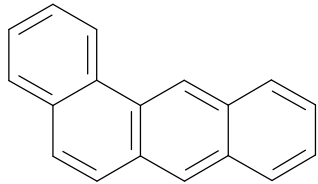

benzo[a]anthracene

Scheme 1 – A selection of PAHs detected by mass spectrometry in the RAE oil fraction.

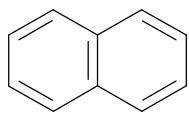

Naphthalene

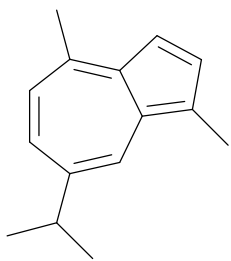

Guaiazulene

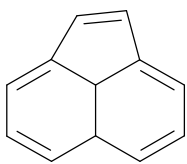

Acenaphthylene

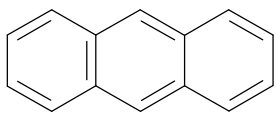

Anthracene

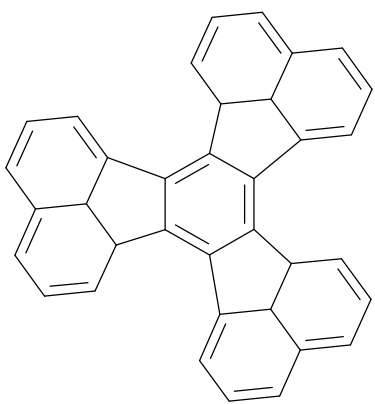

Decacyclene

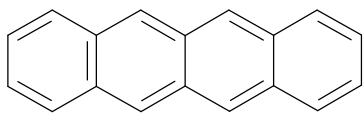

Tetracene

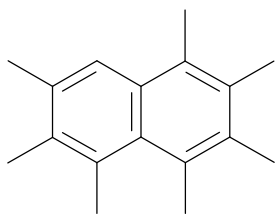

Heptamethylnaphthalene

Scheme 2 – The PAH species carbonized at 700°C under inert atmosphere.

Tables

| Stronger UIE | Weaker UIE | DAE | Coal +250°C | Coal +50°C | Coal -180°C | RAE +50°C | RAE -178°C | Assignments |
|---|---|---|---|---|---|---|---|---|
| | | 2.96 | | | 2.96 | 2.91 | 2.91 | |
| 3.3 | | 3.30 | 3.29 | 3.29 | 3.30 | | 3.37 | Aromatic C-H stretching |
| | 3.4 | | 3.43 | 3.42 | 3.42 | | 3.44 | Aliphatic C-H stretching |
| | 3.5 | | 3.50 | 3.50 | 3.51 | | 3.51 | Aliphatic C-H stretching |
| | 5.25 | | | | | | | |
| | 5.75 | | | | | | | |
| | 6.00 | | | | | | | |
| 6.2 | | 6.25 | 6.27 | 6.26 | 6.26 | 6.26 | 6.25 | Polycyclic aromatic C-C stretching |
| | | | | | | 6.86 | 6.84 | |
| | 6.9 | 6.86 | 6.94 | 6.94 | 6.95 | 7.27 | 7.27 | Aromatic C-C stretching |
| | | 7.07 sh | | | | | | |
| | 7.2-7.5 | 7.26 | 7.27 | 7.28 | 7.29 | | | aliphatic C-H deformation |
| | | | 7.43 | | 7.39 | | | |
| 7.7 | | 7.62 | 7.54 | | 7.54 | 7.64 | 7.65 | aliphatic C-H deformation |
| | | 8.00 | 7.95 | 7.92 | 7.92 | | | aliphatic C-H deformation |
| | | | | | 8.42 | | | aliphatic C-H deformation |
| 8.6 | | 8.56 | 8.50 | 8.62 | 8.55 | 8.60 | 8.54 | aliphatic C-H deformation |
| | | | | | 8.65 | | 8.63 | |
| | | | 9.14 | | 9.06 | | 9.40 | substituted benzene ring frequency |
| | | 9.69 | 9.71 | 9.68 | 9.67 | 9.73 | 9.67 | substituted benzene ring frequency |
| | | | 9.95 sh | | 9.86 sh | | 9.86 | |
| | 10.5 | 10.37 | 10.61 | | 10.62 | 10.36 | 10.36 | aromatic out of plane CH bending |
| | | | | | 10.91 | | | isolated aromatic C-H bending |
| 11.2 | | | | | | | 11.25 | isolated aromatic C-H bending |
| | | 11.51 | 11.60 | 11.59 | 11.53 11.65 | 11.52 | 11.53 | aromatic C-H bending two adjacent C-H |
| | | | | | 12.04 sh | 12.30 | 12.25 | aromatic C-H bending two adjacent C-H |
| | | 12.34 | 12.44 | 12.37 | 12.42 | | | aromatic C-H bending three adjacent C-H |
| 12.7 | | 12.79 | | | | | | aromatic C-H bending three adjacent C-H |
| | | 13.13 | | | 13.21 sh | | | aromatic C-H bending three adjacent C-H |
| | 13.5 | 13.37 | 13.46 sh | 13.41 sh | 13.47 sh | 13.34 | 13.34 | aromatic C-H bending four adjacent C-H |
| | | 13.75 | | | | 13.77 | 13.66 | aromatic C-H bending four adjacent C-H |
| | 14.2 | | | | 14.20 sh | | 13.82 | aromatic C-H bending four adjacent C-H |
| | | | | | 14.37 sh | | 14.26 | mono, meta and 1,3,5 |

**TABLE 1 - UIE IN COMPARISON TO THE INFRARED BANDS OF COAL AND PETROLEUM FRACTIONS DAE AND RAE (WAVELENGTHS IN μm)**

| | | | | | | | |
|---|---|---|---|---|---|---|---|
| | | | 15.10 | | 15.07 | | |
| | | | | | 15.49 | | |
| | | | 15.97 | | 15.95 | | |
| 16.4 | | | | | 16.50 | | |
| | | | 16.93 | | | | |
| | | | | | | 17.25 | 17.33 |
| | | | 17.47 | | 17.59 | | |
| | | | 17.90 | 17.85 | 17.95 | | |
| | | | 18.48 | | 18.45 | | |
| | 18.9 | | 18.87 | | | | |
| | | | 19.50 sh | | 19.52 sh | | |
| | | | 20.19 | | | | |
| | | 20.99 | 20.94 sh | 21.10 sh | 20.89 sh | | 21.01 |
| | | | 21.34 | | 21.26 | | |
| | | | | | 22.19 | | 22.09 |
| | | | 23.04 | | 23.04 | 23.04 | 23.19 |
| | | 23.32 | | 23.39 | 23.58 sh | | 23.66 |
| | | | 24.24 | | 24.10 | | 24.57 |
| | | | 24.81 | | 24.92 | | |

note: sh = shoulder (i.e side band partially covered by the main band)

| TABLE 2 - SUMMARY OF THE INFRARED BANDS OF CARBONIZED ACENES AND OTHER SUBSTRATES (WAVELENTGHTS IN µm) | | | | | | | | | |
|---|---|---|---|---|---|---|---|---|---|
| Stronger UIE | Weaker UIE | Carbonized Naphthalene | Carbonized Acenaphthylene | Carbonized Anthracene | Carbonized Tetracene | Carbonized Pentacene | Carbonized Guaizulene | Carbonized Decacyclene | Carbonized Hexamethylnaphthalene |
| 3.3 | | 3.30 | 3.30 | 3.31 | 3.31 | | 3.30 | 3.29 | 3.31 |
| | 3.4 | 3.42 | | | | | | | |
| | 3.5 | | | | | | | | |
| | 5.25 | | | 5.27 | | | | | |
| | 5.75 | | | 5.78 | | | | | |
| | 6.0 | | | | 6.03 | | | | |
| 6.2 | | | | 6.25 | | | | | |
| | | 6.35 | 6.34 | | 6.34 | | 6.35 | 6.34 | 6.40 |
| | 6.9 | | | 7.00 | 7.00 | | 7.05 | 7.09 | |
| | | | 7.10 | | | | | | |
| | 7.2-7.5 | | | | | | 7.39 | 7.23 | 7.28 |
| 7.7 | | | | | | | | | |
| | | | | | | | | | 7.89 |
| 8.6 | | 8.67 | 8.67 | | 8.68 | | 8.59 | | |
| | | | | 8.75 | | | | | 8.72 |
| | | | | | | | | 9.15 | |
| | 10.5 | | 10.46 | 10.59 | 10.58 | 10.58 | 10.55 | | |
| 11.2 | | | | 11.21sh | | 11.14sh | | | |
| | | 11.55 | 11.55 | 11.50 | 11.54 | 11.51 | 11.55 | 11.59 | 11.55 |
| | | | | | | 11.98 | 11.99 | | |
| | | 12.29 | 12.25 | | | | | | 12.28 |
| | | | | 12.41 | 12.47 | 12.35 | 12.40 | | 12.36 |
| | | 12.60 sh | | | | | | 12.57sh | |
| 12.7 | | 12.75 sh | | | | | | | |
| | | 13.20 | 13.17 | | | | | 13.20 | |
| | 13.5 | | 13.51 sh | 13.51 | 13.52 | 13.53 | 13.48 | | 13.42 |
| | 14.2 | | | | | | | | |
| | | 15.03 | 14.94 | | | | | | |
| | | | | | | 15.48 | | | |
| | | 15.30 | | | | | | | |
| 16.4 | | 16.35 | 16.20 | | | 16.37 | | 16.33 | |
| | | 17.08 | | | | | | | |
| | | | | | 17.27 | | | | |
| | | | | | | 17.43 | | | |
| | | | | | | | | 18.31 | |
| | 18.9 | | | | | | | | |
| | | 19.14 | | | | | | | |
| | | | | | | 20.05 | | | |
| | | 20.85 | | | | | | 21.06 | |

| | | 21.89 | | 21.56 | 21.68 | 21.53 | | | |
| | | | | | | 22.63 | | | |

note: sh = shoulder (i.e side band partially covered by the main band)